\documentclass[twocolumn,showpacs,preprintnumbers,amsmath,amssymb]{revtex4}
\usepackage{graphicx}
\usepackage{dcolumn}
\usepackage{bm}

\newcommand {\be}{\begin{equation}}
\newcommand {\ee}{\end{equation}}
\newcommand {\bea}{\begin{eqnarray}}

\newcommand {\eea}{\end{eqnarray}}

\newcommand {\La}{$\Lambda(1405)$}
\newcommand {\kp}{$K^{-}p$}

\begin{document}

\title{Theoretical analysis of 
$\Lambda(1405) \rightarrow (\Sigma\pi)^0$ mass spectra\\
 produced in $p+p \rightarrow p  + \Lambda(1405) +  K^+$ reactions
}
 
\author{Maryam Hassanvand$^{1,2}$, Seyed Zafarollah Kalantari$^{2}$, 
Yoshinori Akaishi$^{1,3}$, and Toshimitsu Yamazaki$^{1,4}$}

\address{$^{1}$ RIKEN, Nishina Center, Wako, Saitama 351-0198, Japan}
\address{$^{2}$ Department of  Physics, Isfahan University of Technology, Isfahan 84156-83111, Iran}
\address{$^{3}$ College of Science and Technology, Nihon University, Funabashi, Chiba 274-8501, Japan}
\address{$^{4}$ Department of Physics, University of Tokyo, Bunkyo-ku, Tokyo 113-0033, Japan}

\thanks{} 
\date{\today; published in PRC 87, 055202 (2013) + Errata}

\begin{abstract}
 We formulated the $\Lambda(1405)$ (abbreviated as $\Lambda^*$) $\rightarrow (\Sigma\pi)^0$ invariant-mass spectra
 produced in $p+p \rightarrow p +  \Lambda^* + K^+$ reactions, in which both the incident channel for a quasi bound $K^-p$ state and its decay process to $(\Sigma \pi)^0$ were taken into account realistically. We calculated $M(\Sigma \pi)$ spectral shapes for various theoretical models for $\Lambda^*$. These asymmetric and skewed shapes were then compared with recent experimental data of HADES, yielding $M(\Lambda^*) = 1405_{-9}^{+11}$MeV/$c^2$ and $\Gamma = 62 \pm 10$ MeV, where the interference effects of the $\bar{K} N$-$\Sigma \pi$ resonance with the $I$ =  0 and 1 $\Sigma \pi$ continuum are considered. The nearly isotropic proton distribution observed in DISTO and HADES is ascribed to a short collision length in the production of $\Lambda^*$, which justifies the high sticking mechanism of $\Lambda^*$ and the participating proton into $K^-pp$.    
\end{abstract}


\maketitle

\section{Introduction}

The {\La} resonance discovered in 1961 \cite{L1405} (called herein $\Lambda^*$) has strangeness $S$ = -1, spin-parity $J^p$ = ($\frac{1}{2}$)$^-$ , and isospin $I$ = 0. It has been interpreted as a quasi bound state of {\kp} embedded in the $\Sigma + \pi$ continuum since Dalitz-Tuan's original prediction \cite{Dalitz59}. In recent years, Akaishi {\it et al.} derived phenomenologically a complex $\bar{K}N$ interaction (called here the {\it AY} interaction) \cite{AY02PRC,YA07PRC,AMY08} based on the mass and width of {\La}, $M = 1405.1_{-1.0}^{+1.3}$ MeV/$c^2$ and $\Gamma = 50 \pm 2$ MeV \cite{Dalitz-Deloff,PDG,PDG12}, $[$the so-called {\La} ansatz$]$. They applied this very attractive interaction to few-nucleon systems involving one and two $\bar{K}$'s, and found nuclear bound states with unusually high nuclear density \cite{AY02PRC,AY02PLB,AY04PLBa,AY04PRC,AY04PLBb}.
On the other hand, 
 a totally different framework with a double-pole structure of {\La} has emerged on the basis of chiral SU(3) dynamics (called here {\it Chiral}), on which {\La} is claimed to consist of two poles around 1420 and 1390 MeV/$c^2$, which are coupled mainly to $\bar{K}N$ and $\Sigma \pi $ channels, respectively \cite{JidoNPA03,Magas05}. Then, the resulting weakly attractive $\bar{K} N$ interaction leads to much shallower $\bar{K}$ bound states \cite{HW08PRC,DHW}. 
 
 Thus, it is vitally important to determine the location of the {\kp} resonance, whether {\La} is located at 1405 MeV/$c^2$ or above 1420 MeV/$c^2$, from experimental data without prejudice. For this purpose we have to treat the {\La} structure with the {\it AY} model and the {\it Chiral} model on equal footing to be compared with experimental data. To resolve this issue, observations of $M(\Sigma\pi)$ spectra associated with resonant formation of $\Lambda^*$ in the stopped-$K^-$ absorption in $^{3,4}$He \cite{Esmaili10}, and also in $d$ \cite{Esmaili11} have been proposed. Whereas old bubble-chamber experiments of stopped $K^-$ in $^4$He \cite{Riley} indicated a preference of $\Lambda(1405)$ over $\Lambda(1420)$ \cite{Esmaili11,PDG12}, a much more precise experiment with a deuteron target is expected at J-PARC \cite{Suzuki}. Alternatively, Jido {\it et al.} \cite{Jido09} proposed an in-flight $K^-$ reaction on $d$, whereas Miyagawa and Haidenbauer \cite{Miyagawa:12} questioned the effectiveness of this method. In any case, old data on the in-flight $K^- + d$ reaction by Braun {\it et al.} \cite{Braun} had a large statistical uncertainty in distinguishing $\Lambda(1420)$ and $\Lambda(1405)$, according to our statistical analysis. Future experiments at J-PARC of both stopped-$K^-$ \cite{Suzuki} and in-flight $K^-$ \cite{E31} on $d$ are expected to give a convincing conclusion. 
  
Recent experiments on high-energy $pp$ collisions have produced important data on the production of $\Lambda(1405)$:  
\begin{eqnarray}
&&p + p \rightarrow p +  \Lambda^* + K^+, \nonumber \\
&& \hspace{2cm}                       \Lambda^* \rightarrow \Sigma^{+,0,-} + \pi^{-,0,+}. \label{eq:main reaction}
\label{eq:Reaction}
\end{eqnarray}
The ANKE experiment at COSY with an incident kinetic energy ($T_p$) of 2.83 GeV by Zychor {\it et al.} \cite{Zychor08} has yielded a ($\Sigma^0 \pi^0)^0$ invariant-mass spectrum. It was analyzed by Geng and Oset \cite{Geng-Oset} based on chiral SU(3) dynamics. They showed that the reaction in the $\Lambda^*$ production region is dominated by the $|T_{21} |^2 k_2$ process, and they claimed that the spectrum develops a  pronounced strength around 1420 MeV/$c^2$, which differs from the 1405 MeV/$c^2$ peak in Hemingway's data \cite{Hemingway} analyzed by the $|T_{22} |^2 k_2$ process \cite{Dalitz-Deloff,PDG} (see also Akaishi {\it et al.} \cite{Akaishi:2010}). This result might have been accepted as evidence for a double-pole structure of $\Lambda^*$ predicted by chiral SU(3) dynamics \cite{JidoNPA03,Magas05}, if the statistics of the data were good enough. The ANKE data were also analyzed by Esmaili {\it et al.} \cite{Esmaili11}, who, on the contrary, showed from a fair statistical comparison between the two models that the data were in more favor of the {\it AY} model, but the statistical significance was not sufficient to conclusively distinguish between {\it Chiral} and {\it AY } models. Thus, new data from HADES of GSI, which have just been published \cite{HADES12a,HADES12b}, are valuable for solving the present controversy.

 In the present paper we formulate the spectral shape of the $(\Sigma\pi)^0$ mass to provide theoretical guides to analyze experimental data of $(\Sigma\pi)^0$ mass spectra from the above reaction. 
We take into account both the formation and the decay processes of {\La} in $pp$ reactions realistically, following our $\bar{K}N-\Sigma\pi$ coupled-channel formalism \cite{AMY08}. In this way, we derive the general form of the spectral function, which is not symmetric, but skewed with respect to the pole position. Then, 
we analyze  $(\Sigma^{+-}\pi^{-+})^{0}$ spectra from HADES at $T_p$ = 3.50 GeV \cite{HADES12b}. 

\section{Formulation}
\subsection{Coupled-channel treatment of $\Lambda^*$}

Our coupled-channel treatment of $\Lambda (1405)$ is described in \cite{AMY08,Esmaili11}. We employ a set of separable potentials with a Yukawa-type form factor,
\begin{eqnarray}
\langle \vec k'_i  | v_{ij}  | \vec k_j \rangle&=&g(\vec k'_i)\,U_{ij}\, g(\vec k_j),\\
g(\vec k) &=& \frac {\Lambda^2} {\Lambda^2 + \vec k^2}, \\
U_{ij}&=&\frac{1}{\pi^2} \frac{\hbar^2}{2 \sqrt{\mu_i \mu_j}} \frac{1}{\Lambda} s_{ij},
\end{eqnarray}
where $i$ $(j) $ stands for the $\bar KN$ channel, 1, or the $\pi \Sigma$ channel, 2, and $\mu_i$ $ (\mu_j)$ is the reduced mass of channel $i$ $ (j)$. Two of the non-dimensional strength parameters, $s_{11}$ and $s_{12}$, with a fixed $s_{22}$ are adjusted so as to reproduce a set of assumed $M$ and $\Gamma$ values for the $\Lambda^*$ pole \cite{AMY08}.
The transition matrices,
\begin{equation}
\langle \vec k'_i |t_{ij} | \vec k_j \rangle = g(\vec k'_i) \, T_{ij} \,g(\vec k_j),
\end{equation}
satisfy
\begin{eqnarray}
T_{ij}&=&U_{ij}+ \sum_l U_{il} \, G_{l} \,T_{lj}, \\
G_l&=& \frac {2\mu_l}{\hbar^2} \int d\vec q ~g(\vec q) \, \frac {1}{k_l^2-q^2+i\epsilon} \, g(\vec q).
\end{eqnarray} 
The solution is given in a matrix form by
\begin{equation}
T=[1-UG]^{-1}U
\end{equation}
with 
\begin{equation}
(UG)_{lj}=-s_{lj}\sqrt{\frac{\mu_j}{\mu_l}} \, \frac{\Lambda^2}{(\Lambda-i \,k_j)^2},
\end{equation}
where $k_j$ is a relative momentum in channel $j$.

Among the matrix elements, $T_{11}$, $T_{12}$, $T_{21}$ and $T_{22}$, the experimentally observable quantities below the $\bar{K} + N$ threshold are $- (1/\pi ) \, {\rm Im} \, T_{11}$, $| T_{21} | ^2 k_2$ and $|T_{22} | ^2 k_2 $, where the second term with $g^2(k_2)\,g^2(k_1)$ is a $\Sigma \pi $ invariant-mass spectrum from the conversion process, $\bar{K} N \rightarrow \Sigma \pi $ (which we call the ``$T_{21}$ invariant mass"). The $T_{21}$ invariant mass coincides with the $\bar{K}N$ missing-mass spectrum in the mass region below the $\bar{K}+ N$ threshold, as denoted by  relation \cite{Esmaili11}, that
\be
{\rm Im} \,T_{11} = |T_{21}|  ^2 \, {\rm Im}\, G_2.
\ee
The third term with $g^4(k_2)$ is a $\Sigma \pi $ invariant-mass spectrum from the scattering process, $\Sigma \pi \longrightarrow \Sigma \pi $ (which we call the ``$T_{22}$ invariant mass").

\subsection{$\Lambda^* \rightarrow (\Sigma \pi)^0$ spectrum shape}

\begin{figure}
 \center\includegraphics[width=7cm]{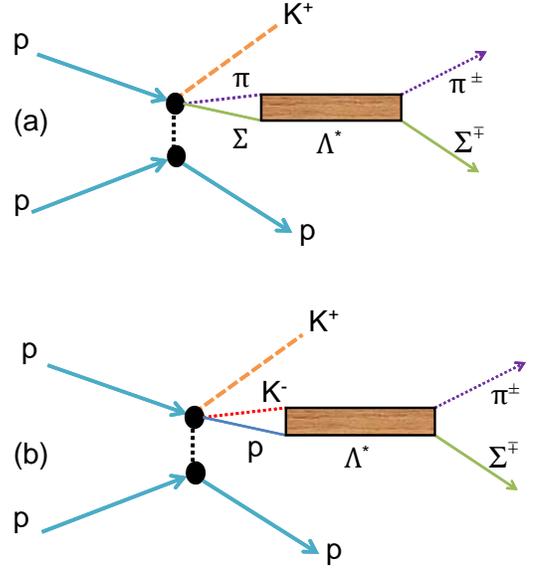}\\
    \caption{(Color online) Feynman diagrams for the $p+p \rightarrow p+K^++\Lambda ^* \rightarrow p+K^+ +(\Sigma \pi )^0$ reaction for (a) the process via $T_{22}$ and (b) the process via $T_{21}$.}\label{fig:fig1}
\end{figure}

\begin{figure}
  \center\includegraphics[width=12cm]{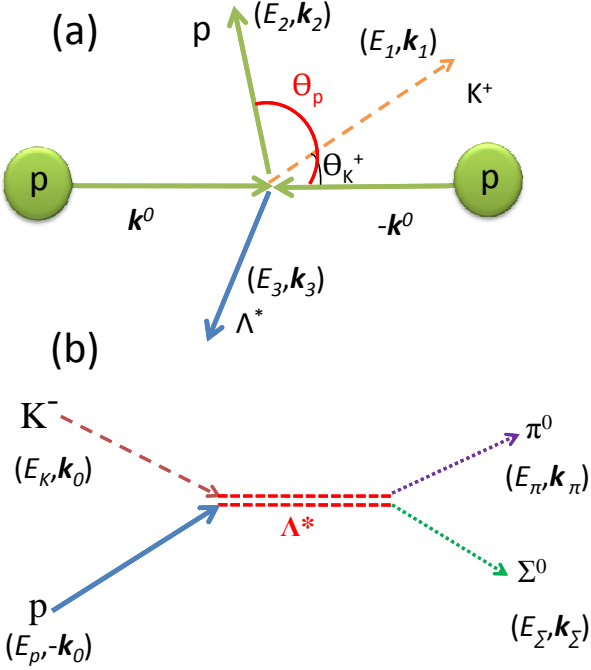}\\
    \caption{(Color online) Kinematical variables in the center of mass of the $p p$ collision for (a) the formation process, $W_{\rm form}$, and (b) the decay channel, $G(x)$. }\label{fig:kinematics}
\end{figure}

The diagram for the reaction Eq.~(\ref{eq:main reaction}) is shown in Fig.~\ref{fig:fig1}.  The decay processes via $T_{21}$ and $T_{22}$ are also given in this figure. The kinematical variables in the c.m. of the $p p$ collision for both the formation and the decay processes are given in Fig.~\ref{fig:kinematics}.

In the present reaction we use $|T_{21}|^2 k_2$ because the incident channel to bring {\La} is $K^- + p$ together with $K^+$ $[$see Fig.~\ref{fig:fig1}(b)$]$. This was also concluded by Geng and Oset \cite{Geng-Oset}, who studied the reaction mechanism in detail. The 
$|T_{22}|^2 k_2$ spectrum would be applicable when $\Sigma$ and $\pi$ mesons are available in the incident channel, as shown in Fig.~\ref{fig:fig1}(a). The $|T_{22}|^2 k_2$ spectrum is characterized by a large tail \cite{Esmaili11} in the higher-mass region up to the kinematical limit, which can in principle be recognizable by an observed spectrum. Experimentally, however, a bump in the upper-tail region may be masked by an ambiguous shape of the continuous background, and may thus be difficult to extract. We may allow a small admixture of $|T_{22}|^2 k_2$ in our likelihood analysis of the experimental data.

The $|T_{21}|^2 k_2$ and $|T_{22}|^2 k_2$ curves of the {\it Chiral} model, as given by Hyodo and Weise \cite{HW08PRC} as well as those of the {\it AY} model, are shown in Fig. 1 (upper) of Ref. \cite{Esmaili11}. They will be compared with the new HADES data at the end of the present paper.


\subsection{Spectral function in the $pp$ reaction: $S(x)$}

Now, we consider the spectrum function of the invariant mass, $S(x)$, in the case of $pp$ reactions. We compose it in the impulse approximation framework by using the incident channel function, $W_{\rm form}(x)$, and the decay channel one, $G(x)$, as follows: 
\be 
S(x) = W_{\rm form}(x) \times G(x),\label{eq:S=WG}
\ee
with 
\begin{equation}
 x = M(\Sigma\pi).
\end{equation}
$G(x)$ is expressed in terms of the $T$ matrices, $T_{22}$ and $T_{21}$, as shown in Figs.~\ref{fig:fig1}-(a) and ~\ref{fig:fig1}-(b). Each function calculated for an assumed $M$ of the $\Lambda^*$ pole is shown in Fig.~\ref{fig:WGS}.

\subsection{Formation process function: $W_{\rm form}$} 

The $\Lambda ^*$ formation from $pp$ collision is calculated in a similar way as was done in \cite{YA07PRC}. We apply an impulse approximation to the formation process of Fig. 1 with a model impulse $t$ matrix, 
\begin{eqnarray}
&&\langle \vec r_{\Lambda^*-p}, \vec r_{(\Lambda^*p)-K^+} | t |\vec r_{p-p} \rangle \nonumber\\
&& = T_0 ~\delta(\vec r_{\Lambda^*-K^+})
 \int d\vec r ~\frac {{\rm exp}(-r/b)}{b^2 r} \delta(\vec r_{\Lambda^*-p}-\vec r) \delta(\vec r_{p-p}-\vec r), \nonumber \\
 &&
\end{eqnarray}
where $\vec r_{a-b}=\vec r_a-\vec r_b$, $T_0$ is a strength parameter, and $b= m_B c/\hbar$ is a range which affects the dependence of the reaction amplitude on the momentum transfer to the adjacent proton in the $pp \rightarrow K^+\Lambda^*p$ process.
Then, the $\Lambda ^*$ formation probability is given as follows:
\bea \label{eq:Wform}
&&W_{\rm form}(x) = \frac {2 \big| T_0 \big| ^2 }{(2\pi )^3 (\hbar c)^6} \frac {E_0}{k_0} \nonumber\\
&&\times \int dE_1  \int d\Omega _1 \, d\Omega _2 \, \Big( \frac {1}{1+b^2 Q^2} \Big) ^2\nonumber \\
&&\times \, k_1 k_2 E_1 E_2 \, \Big[ 1+ \frac{E_2}{E_3} \Big(1+ \frac {k_1}{k_2} \, {\rm cos}\theta _{pK^+} \Big) \Big] ^{-1} \label{Wform}
\eea
where $E_0$ and $k_0$ are the initial energy and momentum in the c.m.  frame, as given by
\be
k_0 = \frac {1}{\hbar} \ \Big[ \frac {1}{2} M_p \,T_p \Big] ^ \frac {1}{2}.
\ee
The other quantities, $k_2, E_2$, and $E_3$, become functions of $x$ due to conservation of momentum and energy, which is applied to all the participating particles to take recoil effects into account. 
Also, $\theta_{pK^+} = ( \theta_p - \theta_{K^+} )$ is the angle between $K^+$ and $p$, $b$ is the range of the $pp$ reaction, and the momentum transfer, $Q$, is
\be
Q = [{k_0}^2 + {k_2}^2 - 2 \, k_0 \, k_2 \, \rm cos \,{\theta _p}]^ \frac {1}{2}. \label{Q}
\ee
As can be seen from the factor, $1/(1 + b^2 Q^2)^2$, a shorter range of $b$ can effectively moderate the strong suppression due to a large momentum transfer, $Q$, in a high-energy $pp$ collision. 

\begin{figure}
  \center\includegraphics[width=8cm]{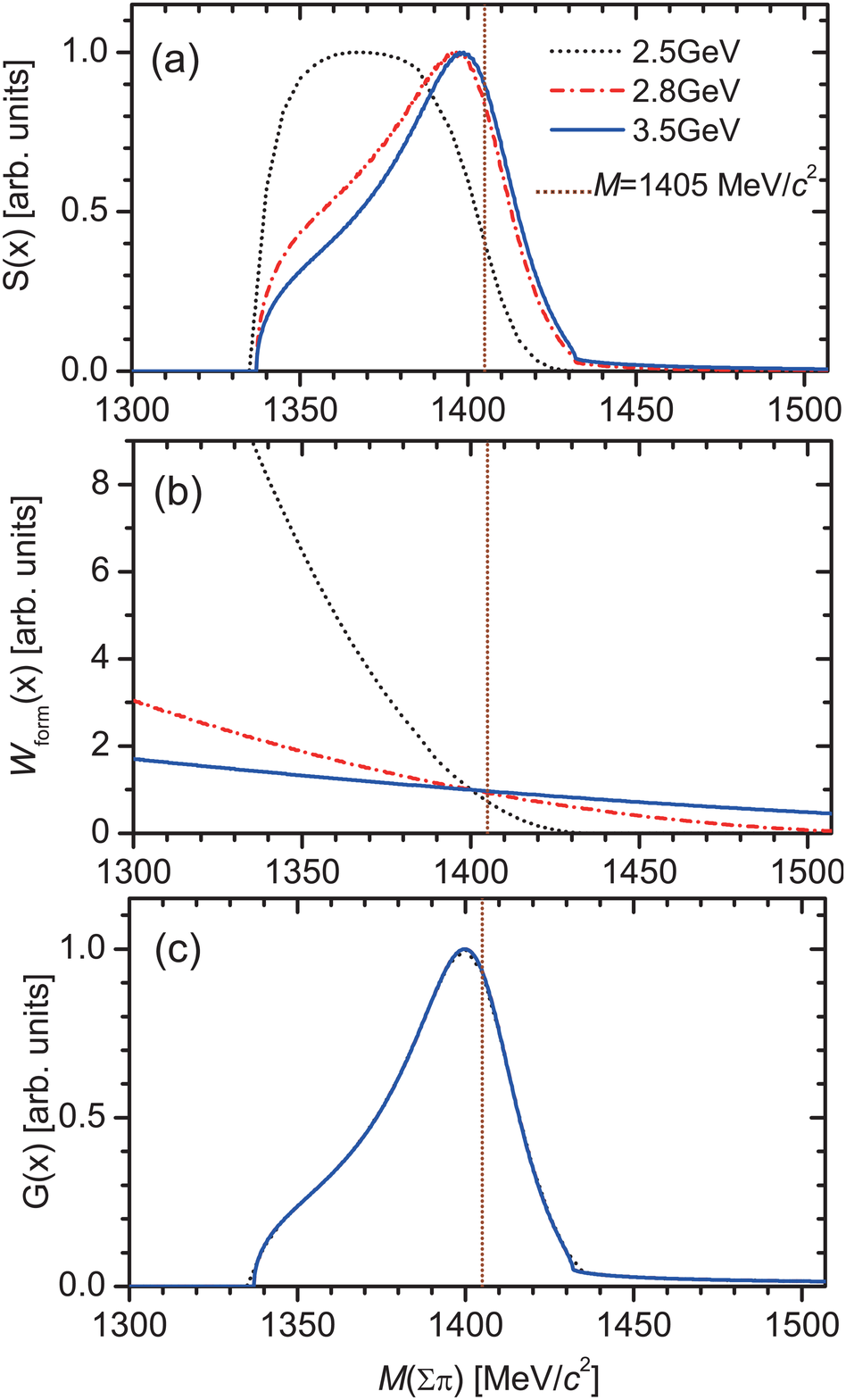}\\
    \caption{(Color online) Normalized spectral functions $S(x)$ (a) composed of the formation-process function $W_{\rm form}$ (b) and the decay-process function $G(x)$ (c) for $T_p$ = 2.50, 2.85 and 3.50 GeV. $m_B = $ 770 MeV/$c^2$ and $(\theta _p, \theta_{pK^+})$ = $(90^\circ ,180^\circ)$. The $M$ value of $\Lambda^*$ is assumed to be 1405 MeV/$c^2$, as indicated by the vertical dashed line.}\label{fig:WGS}
\end{figure}

Figure~\ref{fig:WGS}(b) shows the behavior of $W_{\rm form} (x)$ for $T_p =$ 2.50, 2.83, and 3.50 GeV, the curves of which are normalized at $x = 1400$ MeV/$c^2$. They have respective kinematical upper limits, which make the mass distribution damp toward the kinematical limit. As a result, the observed spectrum shape, $S(x)$, changes, as demonstrated in Fig.~\ref{fig:WGS}(a), whereas $G(x)$ is independent of $T_p$.

\begin{figure}
  \center\includegraphics[width=8cm]{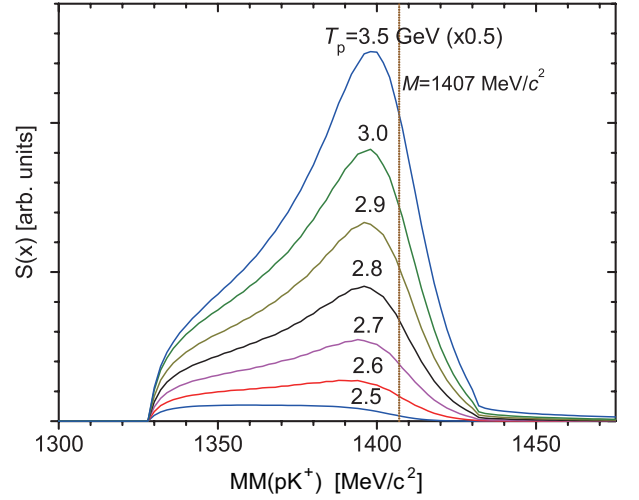}\\
    \caption{(Color online) Incident energy dependence of the absolute values of the spectral function at $m_B = $ 770 MeV/$c^2$ and $(\theta _p, \theta_{pK^+})$ = $(90^\circ ,180^\circ)$. }\label{fig:INCE}
\end{figure}  

\subsection{Decay process function: $G(x)$}

The decay rate of {\La} to ($\Sigma \pi $)$^0$ is calculated by taking into account the emitted $\Sigma$ and $\pi$ particles realistically, following the generalized optical potential formalism in Feshbach theory \cite{Feshbach58}, given by Akaishi {\it et al.} \cite{AMY08,Akaishi:2010}. The decay function, $G(x)$, is not simply a Lorentzian, but is skewed because the kinematic freedom of the decay particles is limited, particularly, when the incident proton energy, $T_p$, decreases and approaches the production threshold. Its general form is given as 
\bea
&& G (x)=\frac{2(2\pi )^5}{\hbar ^2 c^2 } \frac {E_ \pi E_ \Sigma}{E_\pi +E_\Sigma} \,{\rm Re} \,[\tilde{k}(x)] \nonumber \\
&& ~~~~~\times  \big| \langle \tilde{k}(x) \big| t \big| \tilde{k_0}(x) \rangle \big| ^2,   \label{Gama}
\eea 
where the relative momenta in the entrance and exit channels of Fig.~\ref{fig:kinematics}(b)  are calculated by 
\be \label{eq:Sigpi}
\tilde{k_0} (x) = \frac {c \, \surd \lambda (x, m_{K}, M_p)}{2\, \hbar\, x}
\ee
and
\be \label{eq:Sigmapi rel momentum}
\tilde{k} (x) = \frac {c \, \surd \lambda (x, m_\pi, M_\Sigma)}{2\, \hbar\, x}
\ee
with 
\bea
&&\lambda (x, m_1, m_2) \equiv (x+m_1+m_2)(x+m_1-m_2) \nonumber \\
&& ~~~~~~ \times (x-m_1+m_2)(x-m_1-m_2).
\eea

 It should be noticed that $\lambda(x, m_K, M_p)$
 becomes negative at around $x = 1400$ MeV/$c^2$, where we must choose a positive Im $\tilde{k}$ on the physical Riemann sheet. This case corresponds to direct excitation of the $\Lambda^*$ quasi bound state from the $pp$ channel.

In the case of {\it AY}, the $T$ matrix is
\be
\langle \tilde{k} \big| t_{21} \big| \tilde{k_0} \rangle = g(\tilde{k}) \, T_{21} \, g(\tilde{k}_0)
\ee
for the $T_{21}$ process, and 
\be
g(\tilde{k})=\frac {\Lambda ^2}{\Lambda ^2 + \tilde{k}^2}
\ee
with $\Lambda = m'_B c/\hbar$, $m'_B$ being the mass of an exchanged boson, and $\tilde{k}$ is the relative momentum of $\Sigma$ and $\pi$. 

The shape of $G(x)$, as given by Eq.(\ref{Gama}), includes the momenta $\tilde{k_0}$ and $\tilde{k}$, which are functions of $T_p$. However, the function $G(x)$ is shown to depend only on the invariant-mass $x$; namely, $G(x)$ is a unique function of $x$ and does not depend on $T_p$. It is bounded by the lower end ($M_l = M_{\Sigma} + m_{\pi}$ = 1328 MeV/$c^2$) and the upper end ($M_u = M_p + m_{K^-}$ = 1432 MeV/$c^2$). 

It is to be noted that the position of the peak in $G(x)$ is significantly lower than the position of the pole ($M =$ 1405 MeV/$c^2$) in $T_{21}$, as assumed here and indicated by the vertical dashed line. Furthermore, the position of the peak (or centroid) of $S(x)$ is lowered due to the formation channel function $W_{\rm form} (x)$.

\section{Numerical results}
In this section we present results from numerical calculations, and we discuss their physical implications. The importance of the present work is to consider both $W_{\rm form}(x)$ and $G(x) $ functions. In most illustrative samples, we applied the {\it AY} model with the Particle Data Group  (PDG) parameters of \cite{PDG}, $M$ = 1407 MeV/$c^2$ and $\Gamma = 50$ MeV. To compare the {\it Chiral} model with the {\it AY} model on equal footing, we also applied the same procedure as above to Hyodo-Weise's $T$ matrices to obtain realistic spectrum shapes $S(x)$.


\begin{figure}
\center
\includegraphics[width=8cm]{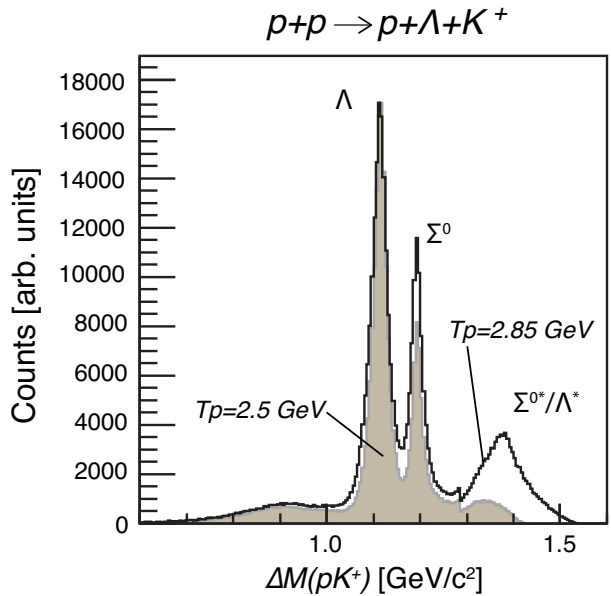}
\caption{(Color online) Experimental spectra of $\Delta M(p K^+)$ in the $pp \rightarrow p \Lambda K^+$ reaction at $T_p = $ 2.50 and 2.85 GeV in DISTO experiments. Taken from \cite{Kienle11}.} \label{fig:threshold}
\end{figure}  

\subsection{Dependence on the incident energy, $T_p$ }
For Eqs. (\ref{eq:S=WG}), (\ref{Wform}), and (\ref{Gama}) again, it is clear that the spectral function depends on the incident proton energy due to the $W_{\rm form}(x)$ function and $G(x)$. 
Figure~\ref{fig:INCE} shows absolute values of spectral functions $S(x)$ for various incident energies ($T_p$) at $m_B = $ 770 MeV/$c^2$ and $(\theta _p, \theta_{pK^+})$ = $(90^\circ ,180^\circ)$. The shape of $S(x)$ is nearly the same, but toward the reaction threshold ($T_p^{\rm thresh} = 2.42$ GeV) not only does the absolute value diminish, but also the spectral shape changes drastically, as shown in Fig.~\ref{fig:WGS}(a) for the normalized spectral functions at $T_p = $ 3.50, 2.83 and 2.50 GeV. The most extreme case is seen at $T_p$ = 2.50 GeV, where the main part of $x > 1400$ MeV/$c^2$ is missing due to the kinematical constraint, and a very skewed component below 1400 MeV/$c^2$ appears. 

\subsection{Behavior near the production threshold of $T_p$}

The above prediction is indeed in good agreement with the observed spectra of DISTO at $T_p$ = 2.50 and 2.85 GeV \cite{Kienle11}, as shown in Fig.~\ref{fig:threshold}. Even in such a very skewed spectrum, one can extract the decay function, $G(x)$, from an observed spectral function by taking the ratio
\be
{\rm DEV}[G(x)] \equiv \frac{S(x)^{\rm obs}}{W_{\rm form}(x)}  \label{eq:DEV}
\ee
using a calculated  $W_{\rm form}$ function. This is a kind of the {\it deviation spectrum method}  introduced in 
stopped-$K^-$ spectroscopy \cite{Esmaili11}.  

\subsection{Angular distribution and correlation}

\begin{figure}
  \center\includegraphics[width=7cm]{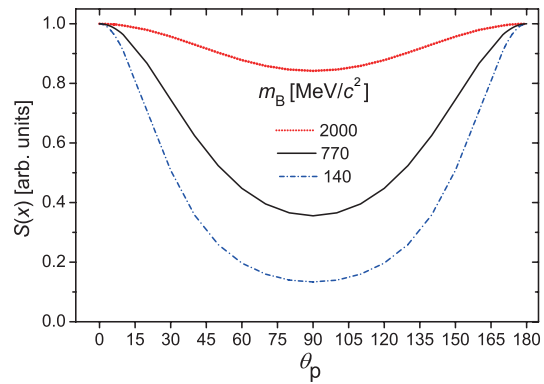}
    \caption{(Color online) Normalized angular distributions of the outgoing proton for different exchanged boson masses, $m_B = $ 2000, 770 and 140 MeV/$c^2$, at $T_p = $ 3.50GeV. }
    \label{fig:symetry}
\end{figure}

The cross section of this reaction has substantial angular dependence (Fig.~\ref{fig:symetry}), but the bound-state peak is distinct at any angle, and we can choose ($ \theta _p, \theta _{pK^+}$) = $(90^\circ ,180^\circ)$, because the cross section is modest and the peak-to-background ratio remains large. The normalized cross sections (spectral shapes) at various angles are found to be nearly the same. Since the two incident protons are indistinguishable, the {\La} formation process is angular symmetric, as shown in Fig.~\ref{fig:symetry}. We can write   
\be
\sigma (\theta _p , \theta_{pK^+}) =  \sigma  (\pi - \theta _p , - \theta_{pK^+})
\ee
for  $\theta _p = 0^\circ - 90 ^\circ$ and 
$\theta_{pK^+} = 0^\circ - 180 ^\circ$. 

\begin{figure}
  \center\includegraphics[width=8cm]{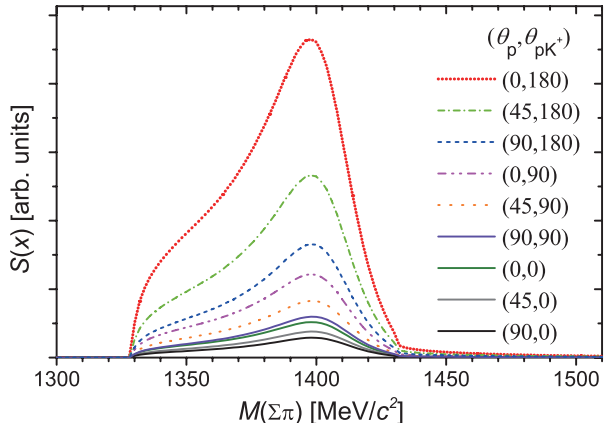}\\
    \caption{(Color online) The spectral functions for various angles, ($\theta _p$, $\theta _{pK^+}$), for $T_p$ = 3.50 GeV and $m_B$ = 770 MeV/$c^2$.}\label{fig:ang-abs}
\end{figure}

According to Eq.~(\ref{Wform}) and Eq.~(\ref{Q}), $W_{\rm form}$, and thus  the spectral function, $S(x)$, are related to the outgoing proton angle, $\theta _p $, and the angle between the outgoing proton and $K^+$, $\theta_{pK^+} $, as shown in Fig.~\ref{fig:ang-abs}. Although these curves look different, the spectrum shape does not depend on the angle. We choose and use $\theta _p$ = 90$^\circ$, $\theta _{pK^+} = 180 ^\circ$ in all of the following calculations. 
\subsection{Dependence on the exchanged boson mass}

Figure~\ref{fig:symetry} shows the normalized angular distributions of the outgoing proton, $\theta _p $, for various masses  of the exchanged boson, $m_B = $ 2000, 770 and 140 MeV/$c^2$, at $T_p = $ 3.50 GeV. The nearly isotropic angular distribution with a large boson mass explains the experimental data of HADES at $T_p = 3.50$ GeV \cite{HADES12a,HADES12b}, which shows that the proton angular distributions together with $\Lambda(1405)$ and $\Lambda(1520)$ are nearly isotropic. A similar behavior is observed in the DISTO data at $T_p$ = 2.85 GeV (see Fig.~\ref{fig:threshold} of the present paper and Refs. \cite{Kienle11,Suzuki12}). Such a short collision length as revealed in the production of 
$\Lambda(1405)$ in the $pp$ reaction is one of the key mechanisms ($\Lambda^*$ doorway) responsible for forming $K^-pp$ from high sticking of $\Lambda^*$ and $p$ \cite{YA07PRC}. On the other hand, it is well known that the  proton emitted in the ordinary $pp \rightarrow p + \Lambda + K^+$ reaction has sharp forward and backward distributions, indicating that the mediating boson is $m_B = m_{\pi}$ \cite{Yamazaki10,Kienle11,Suzuki12}.


\section{$\chi ^2$ fitting of HADES data}

\subsection{HADES data}

In this section we analyze the recent HADES data for charged final states of $\Sigma^- \pi^+$ and $\Sigma^+ \pi^-$ in a $pp$ collision at $T_p$ = 3.50 GeV. The data we use are the missing-mass spectra, $MM(p K^+)$, deduced by the HADES group, as given in Fig. 1 of \cite{HADES12b}, which are corrected for acceptance and efficiency of the detector system. They are expressed as  
\be
Y(x) = Y_{\Lambda^ *} (x) + Y_{\Sigma^ *}(x) + Y_{\Lambda 1520}(x) + Y_{\rm NonRes}(x),
\ee
with $Y_{\Lambda^ *}$ for $\Lambda^*$, $Y_{\Sigma^ *}$ for $\Sigma (1385)$, $Y_{\Lambda 1520}$ for $\Lambda(1520)$, and $Y_{\rm NonRes}$ for the non resonant continuum. The HADES group decomposed the experimental data, $Y(x)$, by the above four components, which were obtained by model simulations, among which the $\Sigma (1385)$ and the $\Lambda(1520)$ components were determined by using the experimental data. The shape of the non-resonant $\Sigma\pi$ continuum was simulated. In their fitting they cautiously excluded the area around 1400 MeV/$c^2$ for $MM(p K^+)$ in order not to bias the finally extracted shape of the $\Lambda^*$ resonance. Then, they found that a simulation of the $\Lambda^*$ region by using a relativistic $s$-wave Breit-Wigner distribution with a width of 50 MeV/$c^2$ and a pole mass of 1385 MeV/$c^2$ can reproduce the experimental data very well, but using instead the nominal mass of 1405 MeV/$c^2$ fails.

This conclusion depends on their assumption of the symmetric Breit-Wigner shape, which is not valid in the case of a broad resonance with adjacent endpoints, $M(\Sigma + \pi)$ and $M(p + K^-)$, as we have seen. Thus, in turn,   
we decided to set up an excess component, $Y_{\Lambda^*} (x)$, by subtracting the given three components from the experimental spectrum $Y(x)$ as
\be
Y_{\Lambda^*} (x) = Y(x) - Y_{\Sigma^*}(x) - Y_{\Lambda 1520}(x) - Y_{\rm NonRes}(x),\label{eq:Y(x)}
\ee
where the statistical errors of $Y(x)$ are inherited to $Y_{\Lambda^*} (x)$. 

\subsection{Interference effects between the $\bar KN$ resonance and the $\Sigma \pi$ continuum}

Before going into the analysis of the HADES data we discuss possible interference effects  between the $\bar KN$ resonance and the $\Sigma \pi$ continuum.

\subsubsection{Interference with the $I$ = 1 $\Sigma \pi$ continuum}

The charge-basis $T$ matrices are related to the isospin-basis $T$ matrices as
\begin{equation}
|T_{\Sigma^+ \pi^-}|^2 \approx \frac{1}{3} |T_{I=0}|^2 + \frac{1}{2} |T_{I=1}|^2 + \sqrt{\frac{2}{3}} {\rm Re}[T_{I=0}^* T_{I=1}], \label{eq;+-}
\end{equation}
\begin{equation}
|T_{\Sigma^- \pi^+}|^2 \approx \frac{1}{3} |T_{I=0}|^2 + \frac{1}{2} |T_{I=1}|^2 - \sqrt{\frac{2}{3}} {\rm Re} [T_{I=0}^* T_{I=1}], \label{eq;-+}
\end{equation}
where $|T_{I=2}|^2$ is neglected.
The HADES $\Sigma^+ \pi^-$ and $\Sigma^- \pi^+$ data show similar behavior: the $\chi^2$ best-fit mass of each of the two spectra is obtained to be very close to one another. This means that the interference term between $I$ = 0 and $I$ = 1 has only a small effect on the resonance spectral shape. Then, we can treat the $I$ = 1 contribution as a part of $Y_{\rm NonRes}$ in the analysis of the $I$ = 0 $\Lambda^*$ resonance, disregarding the interference especially for the sum of the $\Sigma^+ \pi^-$ and $\Sigma^- \pi^+$ data.

\subsubsection{Interference with $I$ = 0 $\Sigma \pi$ continuum}

$\Lambda(1405)$ ( = $\Lambda^*$) is a $I$ = 0 $L$ = 0 $\bar KN$ resonance state  coupled with the $I$ = 0 $L$ = 0 $\Sigma \pi$ continuum. Our theoretical spectrum curves in Fig.~\ref{fig:HKAY+HW} already include the $\bar KN$ threshold effect and also the interference effect with the $I$ = 0 $L$ = 0  $\Sigma \pi$ continuum, because we have solved a $\bar KN$-$\Sigma \pi$ coupled-channel $T$-matrix equation. Thanks to the separation of $Y_{\rm NonRes}$ by the HADES group we need not calculate contributions from the $I$ = 0 $L \geq 1$ $\Sigma \pi$ continuum and $I$ = 1 all $L$ $\Sigma \pi$ continuum, which cause no interference to the $I$ = 0 $L$ = 0 $\Lambda^*$ resonance and therefore can be treated as $Y_{\rm NonRes}$: this is a great advantage of the HADES data for extracting the resonance-pole parameters, the mass and the width of $\Lambda^*$. 

Now we estimate the effect of the $\bar KN$ threshold and the effect of interference with the $I$ = 0 $L$ = 0 $\Sigma \pi$ continuum. By fixing the mass of $\Lambda^*$ to be 1405 MeV/$c^2$, we change AMY's interaction strengths, $s_{11}, s_{12}$ = $s_{21}$, so as to reproduce a given width range of 10 - 70 MeV. The obtained mass spectra are discussed below. 

\begin{figure}[h]
\centering
  \includegraphics[width=8.5cm]{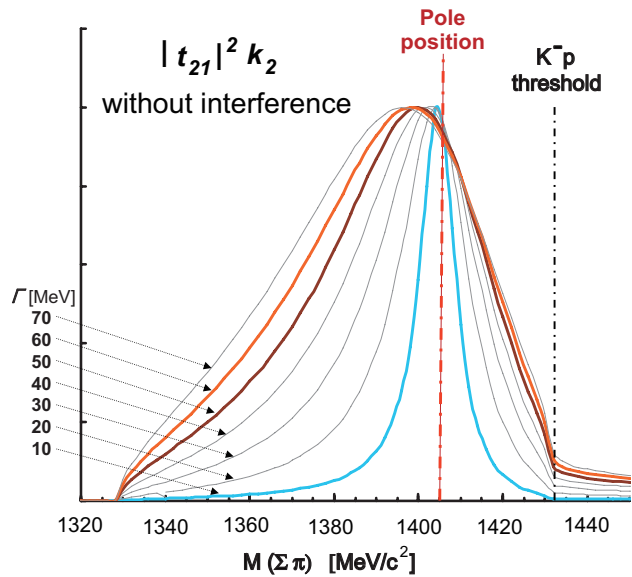}
  \caption{\label{Fig8}Transition mass spectrum, $|t_{21}|^2k_2$, including the $\bar KN$ threshold effect. All the heights are normalized to a same value.}
\end{figure}

\begin{figure}[h]
\centering
  \includegraphics[width=8.5cm]{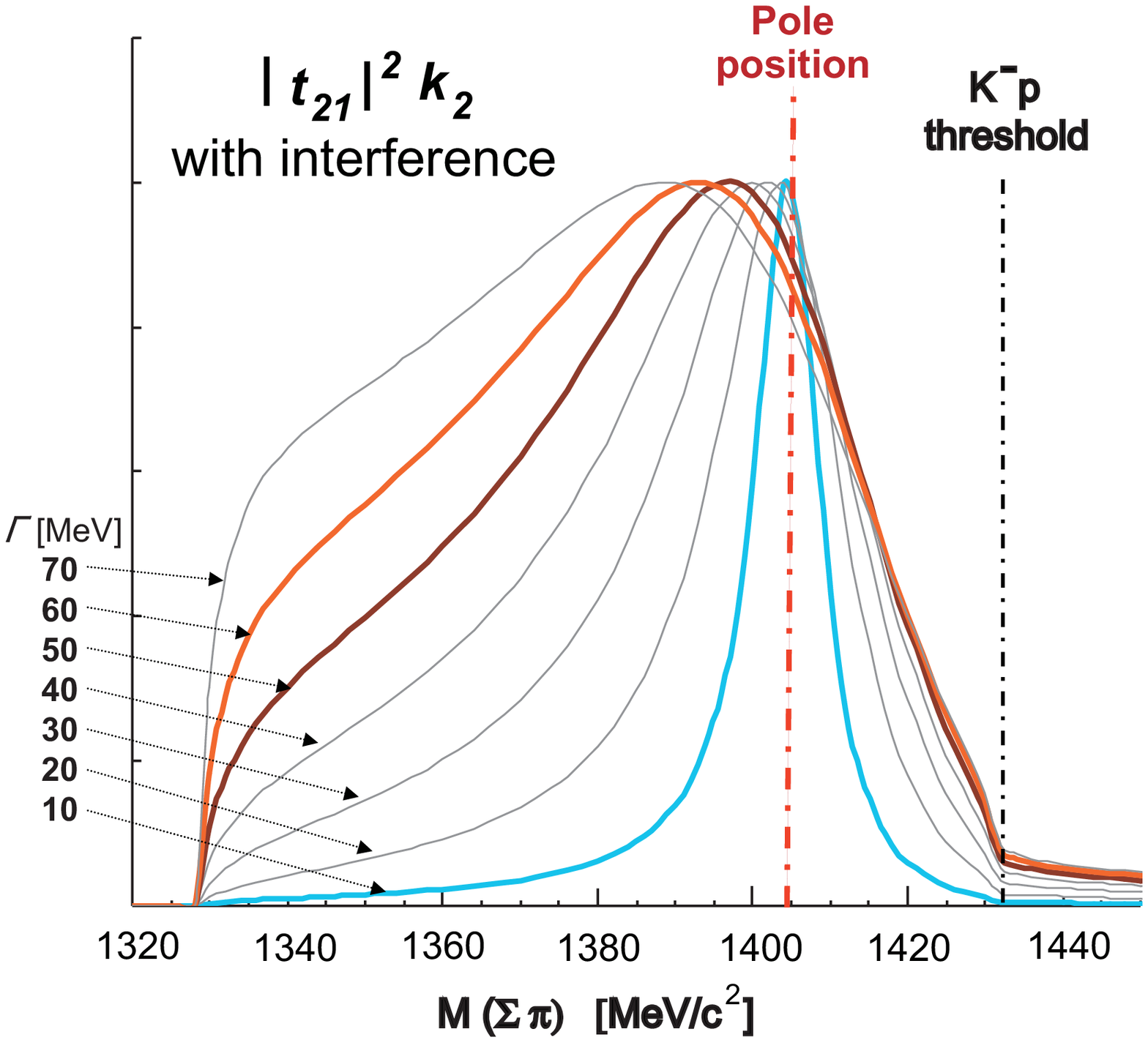}
  \caption{\label{Fig9} Transition mass spectrum, $|t_{21}|^2k_2$, including both the $\bar KN$ threshold effect and the interference effect with the $I$ = 0 $L$ = 0 $\Sigma \pi$ continuum. All the heights are normalized to a same value.}
\end{figure}

Figure~\ref{Fig8} shows the $\bar KN$ threshold effect on the $\Sigma \pi$ invariant mass spectrum, $|t_{21}|^2k_2$, where the interference effect is suppressed by putting $s_{22}=0$. When the width is narrow enough, the spectrum is almost symmetric with a peak close to the pole position. When the width becomes wide, the peak position is lowered from the pole position and the spectrum shape is skewed: this is the $\bar KN$ threshold effect on the spectrum. 
Figure~\ref{Fig9} shows results when the interference effect with the $I$ = 0 $L$ = 0 $\Sigma \pi$ continuum is switched on. The interference effect is not so large for the transition mass spectrum, $|t_{21}|^2k_2$, since the entrance channel to form $\Lambda^*$ has no $\Sigma \pi$ continuum component. 

\begin{figure}[h]
\centering
  \includegraphics[width=8.5cm]{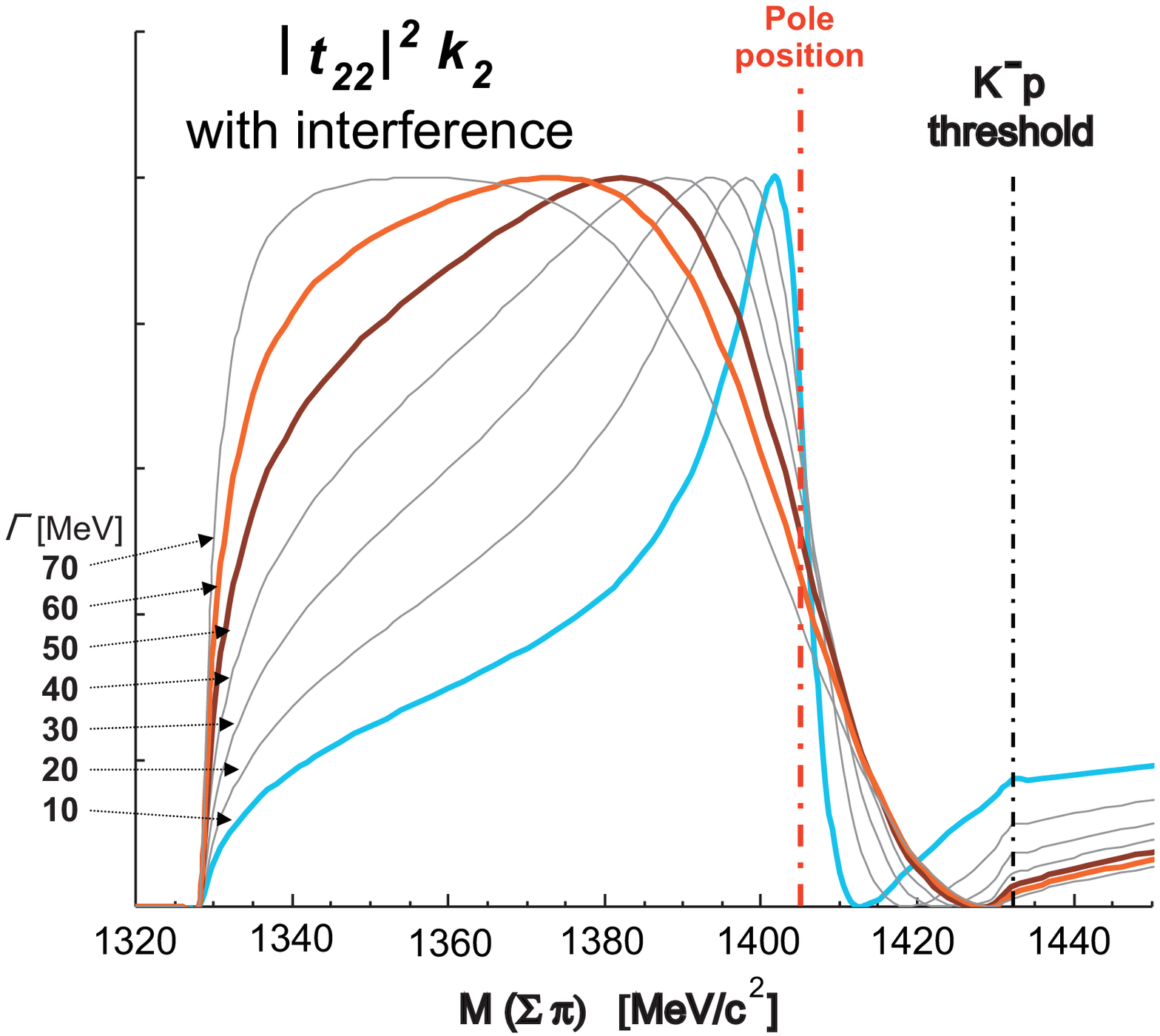}
  \caption{\label{Fig10} Conventional mass spectrum, $|t_{22}|^2k_2$, including both the $\bar KN$ threshold effect and the interference effect with the $I$ = 0 $L$ = 0 $\Sigma \pi$ continuum. All the heights are normalized to a same value.}
\end{figure}

\begin{figure*}
\center\includegraphics[width=13cm]{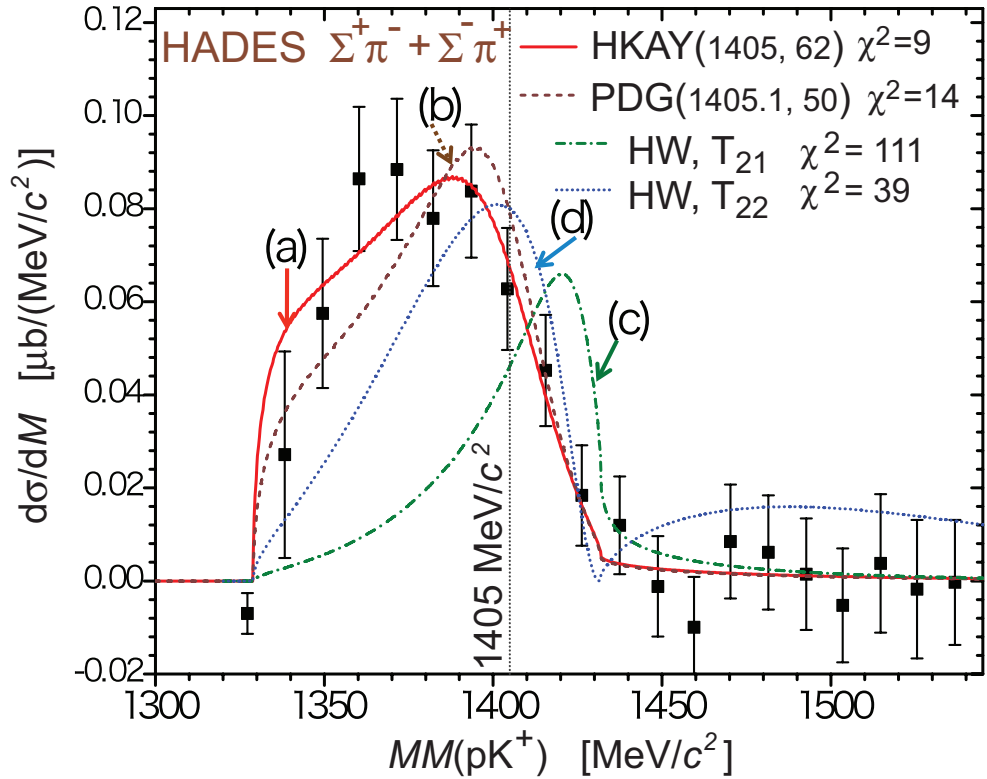}
    \caption{(Color online) Comparison of HADES data ($\Sigma^+ \pi ^- + \Sigma^- \pi ^+$, closed squares) at $T_p = $ 3.50 GeV \cite{HADES12b} with best-fit theoretical spectral functions $S(x)$. (a) Best-fit HKAY curves (with $\chi^2$ = 9.5, $M = 1405_{-9}^{+11}$ MeV/$c^2$, and $\Gamma = 62 \pm 10$ MeV). (b) {\it AY} model with the PDG parameters (with $\chi^2$ = 14, $M = 1405.1 _{-1.0}^{+1.3}$ MeV/$c^2$, and $\Gamma = 50$ MeV \cite{PDG12}). The {\it Chiral} model using HW's $T_{21}$ $[$with $\chi^2$ = 111, (c)$]$ and $T_{22}$ $[$with $\chi^2$ = 40, (d)$]$.   
 }\label{fig:HKAY+HW}
\end{figure*}

On the other hand, Fig.~\ref{Fig10} shows results of the conventional mass spectrum, $|t_{22}|^2k_2$, including the interference effect with the $I$ = 0 $L$ = 0 $\Sigma \pi$ continuum. The interference effect is rather large, since the entrance going to $\Lambda^*$ consists of just $\Sigma \pi$ continuum components, which make the resonance shape deform appreciably. The peak shift comes almost from the interference with the $I$ = 0 $L$ = 0 $\Sigma \pi$ continuum, as seen from an inflection at the pole position and a succeeding interference minimum (see Fig. 8(b) of \cite{Morimatsu88}). The CLAS data \cite{CLAS2013} seem to be a case of $|t_{22}|^2 \, k_2$ where the coupling with the $\Sigma \pi$ continuum becomes significant. The interference between $I=0$ and $I=1$ $\Sigma \pi$ amplitudes gives rise to a strong charge dependence of $\Sigma^+ \pi^-$, $\Sigma^0 \pi^0$, and $\Sigma^- \pi^+$ mass spectra.

The HADES data are well fitted with the transition mass spectrum, $|t_{21}|^2k_2$, as seen from the resemblance between $\Gamma$ = 60 or 50 MeV curves of Fig.~\ref{Fig9} and (a) or (b) of Fig.~\ref{fig:HKAY+HW}. It is noted that the peak shift takes place mainly due to the $\bar KN$ threshold effect in this case.

\begin{figure}
  \center\includegraphics[width=8cm]{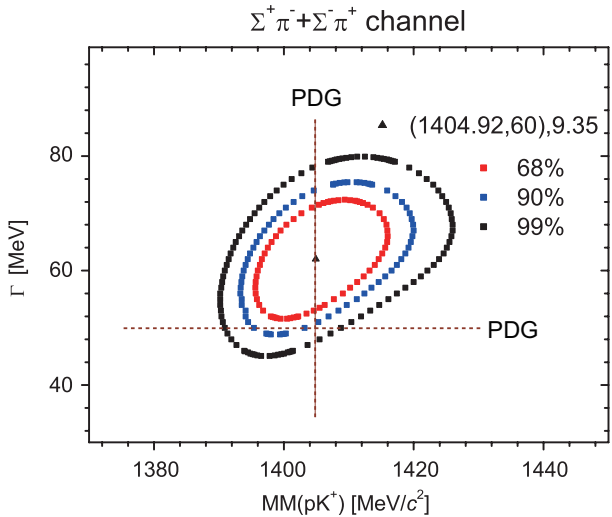} 
    \caption{(Color online) Confidence level contours from $\chi^2$ fitting of the HADES data of $\Sigma^+\pi^- + \Sigma^-\pi^+$ at $T_p$ = 3.50 GeV. The PDG values are also shown.}\label{fig:contour}
\end{figure} 

\subsection{Deduced mass and width}

The HADES spectra, as given in Fig. 1 of \cite{HADES12b}, indicate that the spectra of the two charged channels are similar to each other, yielding nearly the same $M$ values. This fact indicates that the $\Sigma \pi$ resonance is formed by nearly pure charged states, $\Sigma^+ \pi^-$ and $\Sigma^- \pi^+$, without isospin mixing. It also justifies the use of $T_{21}$ for the analysis of $M(\Sigma \pi)$ in the case of $pp$ reactions. On the other hand, the statistical fluctuation of each charged-channel spectrum is rather large. Thus, for the final analysis we use the sum data of HADES ($\Sigma^+ \pi^- + \Sigma^- \pi^+$), which is presented in Fig. 1(c) of \cite{HADES12b}. Keeping the last three components of Eq. (\ref{eq:Y(x)}) fixed, we fit the experimental data of $Y_{\Lambda^ *} (x)$ with $n =$ 21 data points in the range of 1300 to 1550 MeV/$c^2$ (closed points with error bars in Fig.~\ref{fig:HKAY+HW}) 
 by assumed theoretical functions $S(x)$.

 Generally, the experimental histogram, $N_i$, $i$ = 1, ..., $n$, with respective statistical errors, $\sigma _i$, is fitted to a theoretical curve, $S(x; M, \Gamma)$, with $x$ = $MM(p K^+)$ involving the mass $M$ and width $\Gamma$ as free parameters by minimizing the $\chi ^2$ value: 
\be
\chi ^2 (M,\Gamma )= \sum^n_{i=1} \left( \frac{N_i - S(x_i; M, \Gamma)}{\sigma_i} \right)^2.
\ee
Figure~\ref{fig:HKAY+HW} shows the results of the $\chi^2$ fitting, where the HADES data ($\Sigma^+ \pi ^- + \Sigma^- \pi ^+$) at $T_p = $ 3.50 GeV \cite{HADES12b} are compared with best-fit theoretical spectral functions, $S(x)$. The present {\it AY} treatment (hereafter called HKAY), with the PDG values ($M = 1405.1 _{-1.0}^{+1.3}$ MeV/$c^2$ and $\Gamma = 50$ MeV \cite{PDG12}) adopted, gives a remarkable fitting with $\chi ^2$ = 11, which is comparable with the statistically expected value, $<\chi^2>_{\rm exp} \sim 19$. On the other hand, the {\it Chiral} model gives much larger $\chi^2$ values of $\sim$111, when $T_{21}$ is chosen, and of 39, when $T_{22}$ is chosen. Another {\it Chiral} model spectrum by Geng and Oset \cite{Geng-Oset} is almost identical to HW's $T_{21}$.  Thus, the chiral models indicate a substantial deviation from the experimental data. 

Furthermore, we can find best-fit values of $(M, \Gamma)$ from drawing confidence contour curves by varying the parameters $(M, \Gamma)$ in a plane. The results are shown in Fig.~\ref{fig:contour}. From this contour mapping we obtain the following best-fit values with 68\% confidence levels (1$\sigma$) errors: 
\bea
&&M = 1405_{-9}^{+11} ~{\rm MeV}/c^2,\\
&&\Gamma = 62 \pm 10 ~{\rm MeV}. 
\eea
The best-fit curves are shown together with the experimental points in Fig.~\ref{fig:HKAY+HW}. The $M$ value thus obtained from the present analysis of the new HADES data confirms the traditional value \cite{PDG,PDG12}. \\

\section{Concluding remarks}

We have presented results of our calculation for the spectral shape of $MM(p K^+)$ in the $pp \rightarrow p \Lambda^* K^+$ reaction based on the $\bar{K} N$-$\Sigma \pi$ coupled-channel treatment. We took into account both the entrance process and the decay process. The formation probability, $W_{\rm form}$, of $\Lambda ^*$ in a $pp$ collision and the decay rate, $G(x)$, to $(\Sigma \pi)^0$ were formulated. The spectral function is given by $S(x) = W_{\rm form} \times G(x)$. It was found to be asymmetric and skewed due to the kinematic limitation imposed by the entrance channel. The peak of $S(x)$ is not located at the pole position. 

With this tool in hand we analyzed the recent HADES data. The interference effects of the $\bar{K} N$-$\Sigma \pi$ resonance with $I$ = 0 and 1 $\Sigma \pi$ continuum are considered. Although the observed spectra of $MM(p K^+)$ appear to show the peak position at around 1385 MeV/$c^2$, the $\chi ^2$ fitting by our theoretical spectral functions provided $M = 1405 _{-9}^{+11}$ MeV/$c^2$. This value is in good agreement with the values obtained from a recent  analysis \cite{Esmaili10} of an old experimental data of stopped-$K^-$ in $^4$He \cite{Riley}, taken up as the updated PDG value ($M = 1405.1_{-1.0}^{+1.3}$ MeV/$c^2$) \cite{PDG12}. On the other hand, the {\it Chiral} model with $M \sim 1420$ MeV/$c^2$ cannot reproduce the experimental data. 

The Faddeev method is suitable for treating final-state interactions of three particles. However, it is difficult to apply this method to the present high-energy $p$-induced processes where so many partial waves are involved. On the other hand, for the low-energy $K^-+d$ reaction R$\acute{\rm e}$vai \cite{Revai2013} succeeded in extracting the $\Lambda(1405)$ resonance structure by using the Faddeev method. We are considering an analysis future data of stopped $K^-$ on $d$, proposed in \cite{Esmaili11,Suzuki}, by fully taking account of final-state interactions in the Faddeev formalism.

The proton angular distribution in $\Lambda^*$ production was also calculated. The isotropic distribution observed in HADES \cite{HADES12b} and DISTO \cite{Kienle11,Suzuki12} were explained by a short-range collision with an intermediate boson mass heavier than the $\rho$ meson mass. This is consistent with the calculated large cross section for the production of $K^-pp$ in $pp$ collisions \cite{YA07PRC}, which has recently been observed in DISTO experiments \cite{Yamazaki10}.\\

\section{Acknowledgments}

This work is supported by a
Grant-in-Aid for Scientific Research from the Ministry of Science, Research, and Technology of Iran and 
by a Grant-in-Aid for Scientific Research from Monbu kagakusho of Japan. One of us (T Y) acknowledges support by the Alexander von Humboldt Foundation, Germany.


\end{document}